\documentclass[aps,twocolumn,superscriptaddress,showpacs,pre]{revtex4}
\usepackage{graphicx}

\begin{document}
\title{Two-parameters electromagnetic hysteresis of a subwavelength nonlinear $\epsilon$-near-zero slab}

\author{A. Ciattoni}
\affiliation{Consiglio Nazionale delle Ricerche, CNR-SPIN 67100 L'Aquila, Italy and Dipartimento di Fisica, Universit\`{a} dell'Aquila, 67100
             L'Aquila, Italy}

\author{C. Rizza}
\affiliation{Dipartimento di Ingegneria Elettrica e dell'Informazione, Universit\`{a} dell'Aquila, 67100 Monteluco di Roio (L'Aquila), Italy}

\author{E. Palange}
\affiliation{Dipartimento di Ingegneria Elettrica e dell'Informazione, Universit\`{a} dell'Aquila, 67100 Monteluco di Roio (L'Aquila), Italy}

\date{\today}

\begin{abstract}
We consider propagation of transverse magnetic electromagnetic waves through a nonlinear metamaterial slab, of subwavelength thickness and very small
average dielectric permittivity, made up of alternating metal and dielectric nonlinear layers. Exploiting the effective-medium approach, we prove
that the output field intensity is a multi-valued function of both the input field intensity and incidence angle so that the transmissivity exhibits
a novel kind of multi-stability and a pronounced two-parameter hysteresis behavior. The predicted two-parameter hysteresis physically stems from the
fact that, since the linear and nonlinear contributions to the overall dielectric response can be comparable, the slab can host more than one
electromagnetic configuration compatible with the incident field and the boundary conditions, regardless of the slab thickness. We investigate slab
transmission also through full wave simulations proving the robustness of the two-parameter hysteresis beyond the effective-medium approximation.
\end{abstract}
\pacs{78.67.Pt, 42.65.Tg}

\maketitle

Hysteresis is probably one of the most intriguing feature a nonlinear system can exhibit both theoretically and for its potential applications for
the designing of memory devices. Optical bistability and related hysteresis have attracted a large research interest in the last decades
\cite{SzokeD,Chennn} since they are the basic ingredients for devising optical memories, logic gates and optical computing devices \cite{Abraha}. The
advent of metamaterials, with the related possibility of engineering the medium electromagnetic response \cite{Smith2}, has triggered a renewal of
interest towards optical bistability \cite{Litch1}. One-dimensional photonic crystals consisting of alternating layers of positive-index and
negative-index materials index \cite{LiZhou} allow, in the presence of a nonlinearity, bistable switching and tunable nonlinear transmission
\cite{Feisee}, omnidirectional bistability \cite{Hegde2} and bistability at very low values of input intensity \cite{AliAbd}. Optical multistability
have also been predicted in the presence of metal-dielectric multilayer structures \cite{Noskov} and it has been shown that, since for these
structures the sign of the effective dielectric constant depends on the field intensity, steplike transmission of light can be achieved
\cite{Husako}. On the other hand, media with very small permittivity have been shown to support the regime where linear and nonlinear polarizations
are comparable \cite{Ciatt1} and this has led to the prediction of a pronounced transmissivity directional hysteresis exhibited by nonlinear
metamaterial slabs with very small linear permittivity \cite{Ciatt2}.

In this Letter, we investigate the electromagnetic transmission of transverse magnetic (TM) waves through a metamaterial slab of subwavelength
thickness made up of alternating metal and dielectric nonlinear Kerr layers stacked along the direction orthogonal to the thickness. The layers
dielectric properties and filling fractions are chosen so as to yield a negative and very small effective linear permittivity. Exploiting the
effective medium approach, we numerically solve Maxwell equations for the problem of reflection and transmission of an inclined incident TM plane
wave and we show that the slab transmissivity is a multi-valued function of both the input field intensity and incidence angle. The predicted
multistability physically stems from the fact that, since linear and nonlinear dielectric polarizations can be comparable \cite{Ciatt1} (extreme
nonlinear regime), the overall nonlinear dielectric response supports more than one field vector configuration compatible with the incoming plane
wave and the boundary conditions, regardless of the slab thickness. Therefore, the proposed multistability fundamentally differs from standard
optical bistability where the multiple field configurations are produced by a nonlinear feedback process. In order to support our results beyond the
effective medium approximation, we perform full wave simulations whose results are in good agreement with those of the effective medium approach thus
proving the robustness of the predicted phenomenology.

\begin{figure}
\includegraphics[width=0.45\textwidth]{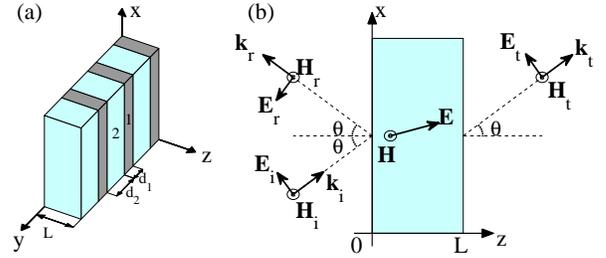}
\caption{(Color online) (a) Geometry of the metamaterial slab. (b) Geometry of the TM incident (i), reflected (r) and transmitted (t) plane waves
together with the TM field within the slab.}
\end{figure}

Consider a TM monochromatic field (with time dependence $\exp(-i \omega t)$) ${\bf E} = E_x(x,z) \hat{\bf e}_x + E_z(x,z) \hat{\bf e}_z$ and ${\bf H}
= H_y(x,z) \hat{\bf e}_y$ propagating through the metamaterial slab, embedded in vacuum, reported in Fig.1(a) of sub-wavelength thickness (along the
$z$-axis) $L = 2/k_0$ ($k_0 = \omega /c$) and consisting of alternating, along the $y$-axis, two nonlinear isotropic and non-magnetic Kerr layers of
thicknesses $d_1$ and $d_2$. We here consider the situation where the layers alternation period is much smaller than the vacuum wavelength (i.e. $d_1
+ d_2 \ll 2 \pi / k_0$) so that, exploiting the effective-medium approach, the slab behaves as an homogeneous medium characterized by the dielectric
response (constitutive relations)
\begin{eqnarray} \label{constituent}
D_x = \epsilon_0 \left\{ \epsilon E_x + \chi \left[\left({\bf E \cdot E^*} \right) E_x + \frac{1}{2} \left({\bf E \cdot E} \right) E_x^* \right] \right\}, \nonumber \\
D_z = \epsilon_0 \left\{ \epsilon E_z + \chi \left[\left({\bf E \cdot E^*} \right) E_z + \frac{1}{2} \left({\bf E \cdot E} \right) E_z^* \right]
\right\},
\end{eqnarray}
where the effective dielectric permittivity $\epsilon = f \epsilon_1 + (1-f) \epsilon_2$ and effective nonlinear susceptibility $\chi = f \chi_1 +
(1-f) \chi_2$ are the averages of their layers counterparts (here $f= d_1/(d_1+d_2)$ is the volume filling fraction of layer $1$) \cite{Ciatt1}.
Therefore, employing both nonlinear negative and positive dielectrics ($Re(\epsilon_1) <0$ and $Re(\epsilon_2) >0$), conditions can be found so that
the effective dielectric permittivity $\epsilon$ has a negative real part (with a very small imaginary part) and very close to zero, and $\chi>0$
(see below for an explicit example) which is the situation we focus on in this Letter.
A TM plane wave $({\bf E}_i,{\bf H}_i)$ incoming from vacuum ($z<0$) and impinging on the slab at $z=0$ with incidence angle $\theta$, produces a TM
reflected plane wave $({\bf E}_r,{\bf H}_r)$ for $z<0$ and a TM transmitted plane wave $({\bf E}_t,{\bf H}_t)$ for $z>L$, as reported in Fig.1(b).
The fields of the three waves are ${\bf E}_s = E_s (\hat{\bf e}_y \times {\bf k}_s/k_0) \exp \left(i {\bf k}_s \cdot {\bf r}\right)$, ${\bf H}_s =
E_s \sqrt{\epsilon_0/\mu_0} \hat{\bf e}_y \exp \left(i {\bf k}_s \cdot {\bf r}\right)$ where $s=(i,r,t)$, $E_i$, $E_r$ and $E_t$ are the amplitudes
of the three waves and ${\bf k}_i={\bf k}_t= k_0 \left(\sin \theta \hat{\bf e}_x + \cos \theta \hat{\bf e}_z \right)$, ${\bf k}_r= k_0 \left(\sin
\theta \hat{\bf e}_x - \cos \theta \hat{\bf e}_z \right)$ are the wave vectors of the three waves. The TM electromagnetic field within the slab is of
the form ${\bf E} = \left[ E_x(z) \hat{\bf e}_x + E_z(z) \hat{\bf e}_z \right] \exp(i k_0 x \sin \theta )$, ${\bf H} = \left[ H_y(z) \hat{\bf e}_y
\right] \exp(i k_0 x \sin \theta )$ so that, Maxwell equations $\nabla \times {\bf E} = i \omega \mu_0 {\bf H}$, $\nabla \times {\bf H} = -i \omega
{\bf D}$  yields the system
\begin{eqnarray} \label{Maxwell}
\frac{d E_x}{dz} &=& i (k_0 \sin \theta) E_z + i \omega \mu_0 H_y, \nonumber \\
\frac{d H_y}{dz} &=& i \omega D_x, \nonumber \\
(k_0 \sin \theta) H_y &=& - \omega D_z,
\end{eqnarray}
where $D_x$ and $D_z$ are given by Eqs.(\ref{constituent}). At the slab input $z=0$ and output $z=L$ faces we require the continuity of the
tangential components of the electric and magnetic fields. In order to evaluate the slab transmissivity $T=|E_t|^2/|E_i|^2$, we specify the amplitude
of the transmitted field $E_t$ at the output face $z=L$ and successively solve Eqs.(\ref{Maxwell}) all the way to the input face at $z=0$. For our
numerical evaluations we have chosen a slab with $\epsilon = -0.05$ and $\chi>0$ and we have discarded the solutions of Eqs.(\ref{Maxwell}) for which
the profile of $E_z$ is discontinuous within the slab $0<z<L$ (a possibility stemming from the fact that the third of Eqs.(\ref{Maxwell}) is a
nonlinear algebraic equation and not a differential equation for $E_z$).
\begin{figure}
\includegraphics[width=0.45\textwidth]{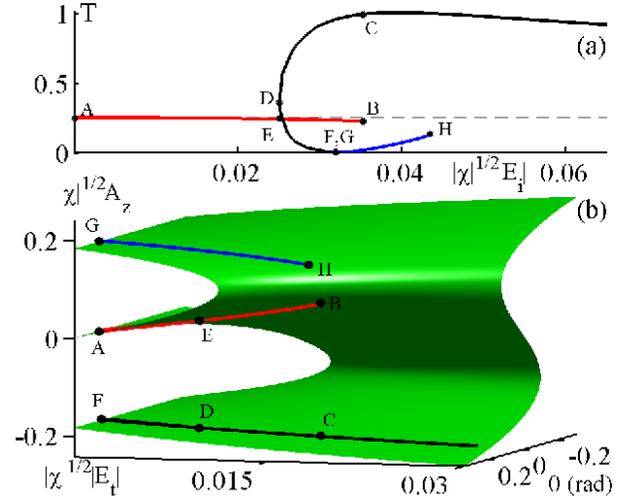}
\caption{(Color online) (a) Nonlinear (solid line) and linear (dashed line) transmissivity $T$ as a function of the normalized input field amplitude
$|\chi|^{1/2} |E_i|$ at the fixed incident angle $\theta=0.17$ rad for a slab with $\epsilon=-0.05$ and $\chi >0$. (b) Surface $|\chi|^{1/2} A_z$
obtained by Eq.(\ref{Ezmatch}), for a slab with $\epsilon=-0.05$ and $\chi >0$.}
\end{figure}
\begin{figure}
\includegraphics[width=0.45\textwidth]{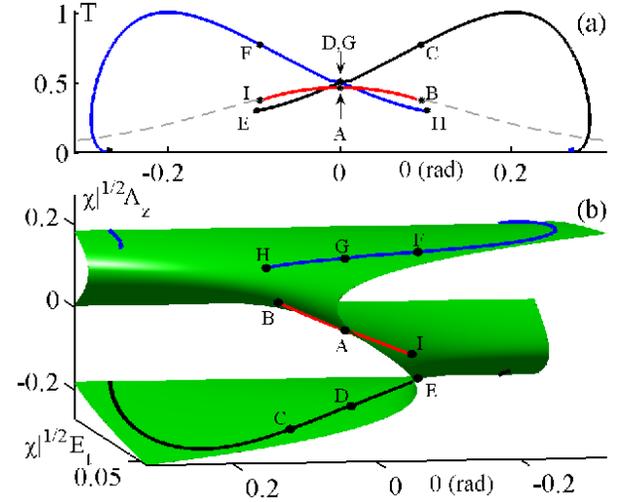}
\caption{(Color online) (a) Nonlinear (solid line) and linear (dashed line) transmissivity $T$ as a function of the incident angle $\theta$ for the
normalized input field amplitude $|\chi|^{1/2} |E_i| =0.05$ for a slab with $\epsilon=-0.05$ and $\chi >0$. (b) Surface $|\chi|^{1/2} A_z$ obtained
by Eq.(\ref{Ezmatch}), for a slab with $\epsilon=-0.05$ and $\chi >0$.}
\end{figure}
In Fig.2(a) we report the evaluated transmissivity $T$ (solid line) as a function of the normalized input field amplitude $|\chi|^{1/2}|E_i|$ at the
fixed incidence angle $\theta=0.17$ rad. In order to evaluate the transmissivity for a given input field amplitude as a function of incidence angle
$\theta$ we have first evaluated a number of amplitudes $E_i$ for various $E_t$ and $\theta$ and subsequently collected all the configurations
corresponding to the same $|E_i|$. In Fig.3(a) we report the resulting slab transmissivity $T$ as a function of $\theta$ for $|\chi|^{1/2} |E_i| =
0.05 $ (solid line). In Fig.2(a) and 3(a) we have also reported, for comparison purposes, the linear ($\chi=0$) slab transmissivity (dashed line)
obtained in the same electromagnetic excitation conditions. Note that both the transmissivities (solid lines) of Fig.2(a) and 3(a) are multi-valued
functions or, more precisely, there are ranges of both the normalized input field amplitude ($0.025 < |\chi|^{1/2} |E_i| < 0.035$) and the incidence
angle ($-0.1$ rad $< \theta < 0.1$ rad) at which the transmissivity is threefold.

In order to discuss the origin of the predicted multistability it is worth noting that the first two of Eqs.(\ref{Maxwell}) are differential
equations describing the spatial dynamics of the fields $E_x$ and $H_y$ whereas the third of Eqs.(\ref{Maxwell}) is a nonlinear algebraic (cubic)
equation yielding a field constraint which allows to evaluate $E_z$ (at least locally) for $E_x$ and $H_y$ prescribed. Therefore, it is evident that,
in general, there are three values of $E_z$ compatible with given $E_x$ and $H_y$ so that, from a physical point of view, the overall nonlinear
response of Eqs.(\ref{constituent}) can generally support different electromagnetic field configurations for a given input intensity. In order to
clarify this point and to discuss the impact of the boundary conditions, note that, from the third of Eqs.(\ref{Maxwell}), it is evident that the
continuity of $H_y$ across the slab boundaries is equivalent to the continuity of $D_z$, so that, after defining the real amplitude $A_z$ through the
relation $E_z(L) = A_z ( E_t/|E_t| ) \exp(2 i \cos \theta)$ and exploiting the second of Eqs.(\ref{constituent}) together with the continuity of
$E_x$, the matching condition $D_z(L^-)=D_z(L^+)$ yields
\begin{equation} \label{Ezmatch}
\left[\epsilon  + \frac{3}{2} \chi \left( |E_t|^2 \cos^2 \theta + A_z^2 \right) \right] A_z+ |E_t| \sin \theta = 0,
\end{equation}
which is a relation that, for a given $E_t$ and $\theta$ yields the possible values of $A_z$. Note that, in the linear regime ($\chi=0$)
Eq.(\ref{Ezmatch}) yields $A_z = - |E_t| \sin \theta / \epsilon$ i.e. the boundary conditions allow only one possible value of $A_z$. The possible
values of $|\chi|^{1/2} A_z$ obtained by Eq.(\ref{Ezmatch}) belong to a surface which is plotted in Figs.2(b) and 3(b), in the case $\epsilon =
-0.05$ and $\chi>0$, and we note that such surface is folded so that there are regions where, for a given pair $|\chi|^{1/2}|E_t|$ and $\theta$,
there are more than one accessible value of $|\chi|^{1/2} A_z$. Therefore the multi-valuedness of the slab transmissivity corresponds to different
fields with the same input intensity and incidence angle having different output longitudinal component $A_z$ belonging to different sheets of the
folded surface. Note that, for $\theta =0$, Eq.(\ref{Ezmatch}) admits the solutions $A_z^{(0)}=0$ and $|\chi|^{1/2} A_z^{(\pm)} = \pm \sqrt{-
\textrm{sign} (\chi)\left(2\epsilon/3 + \chi |E_t|^2 \right)}$ so that the surface can be folded only if $\textrm{sign} (\epsilon \chi) =-1$ and its
upper and lower sheets are such that $\chi A_z^2 \sim |\epsilon|$. This implies that, as opposite to the situation we are considering in this Letter
where $|\epsilon| \ll 1$, in standard materials where $\epsilon$ is of the order of unity, the required nonlinearity would be so large to effectively
forbid the folding of the surface (and the consequent feasibility of the directional multi-stability).

\begin{figure}
\includegraphics[width=0.45\textwidth]{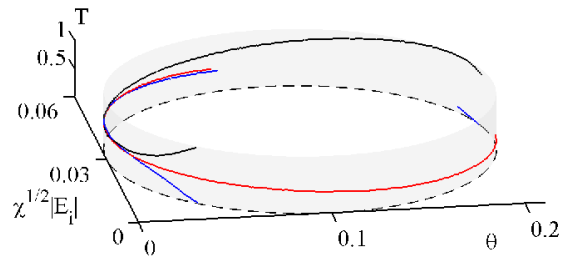}
\caption{(Color online) Transmissivity hysteresis (solid lines) obtained by varying $\theta$ and $\chi^{1/2} |E_i|$ along a closed path (reported as
a dashed line).}
\end{figure}

The predicted multistability is accompanied by a two-parameters transmissivity hysteresis. Note that each of the points of the curves (solid lines)
of Figs.2(a) and 3(a) corresponds to an electromagnetic configuration within the slab and hence these curves have an image on the surface
$|\chi|^{1/2} A_z$ which are reported in Figs. 2(b) and 3(c) as solid lines as well. Suppose to switch on the incident plane wave with $\theta =
0.17$ rad and with an input intensity very small so that the state A of Fig.2(a) and 2(b) is excited since before the illumination the slab was not
polarized. By increasing the input intensity and holding the incidence angle fixed, the transmissivity follows the curve joining the point A and B of
Fig.2(a)
 while the electromagnetic state continuously varies as in Fig. 2(b). From the state B, if the intensity is further increased, it is evident that the
electromagnetic state undergoes a sudden jump to the state C (see Fig.2(b)) on the lower surface sheet since there is no allowed state continuously
joined to B. As a consequence, the transmissivity undergoes a sudden jump to a higher value (see Fig.2(a)). If now, starting from the state C, the
input intensity is decreased, the electromagnetic state varies along the curve from C to D of Fig.2(b) whereas the transmissivity in Fig.2(a) assume
the values along the curve from C to D which are different from those attained along the forward path. At the state D, if the intensity is further
decreased, the state undergoes a jump to the state E of Fig.2(b), belonging to the central surface sheet, since the states from D to F of Fig.2(b)
have, according to Fig.2(a), higher input intensities. An analogous but different hysteretic behavior is obtained if the input intensity is held
fixed and the incidence angle is varied, as depicted in Fig.3. Since the presented multistability depends on two parameters, one can vary both the
incidence angle and intensity on a prescribed path to obtain nontrivial hysteresis behaviors, as reported in Fig.4 where $\theta$ and $\chi^{1/2}
|E_i|$ are varied along a closed path.

\begin{figure}
\includegraphics[width=0.45\textwidth]{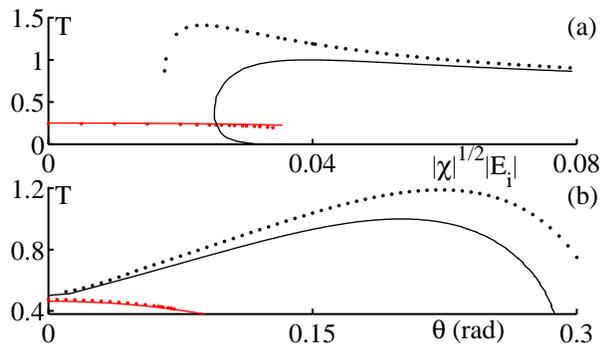}
\caption{(Color online) Comparison between the slab transmissivities evaluated through full-wave simulations (dotted lines) and those of Fig.2(a) and
Fig.3(a) (solid lines), reported in panel (a) and (b), respectively}
\end{figure}

\begin{figure}
\includegraphics[width=0.45\textwidth]{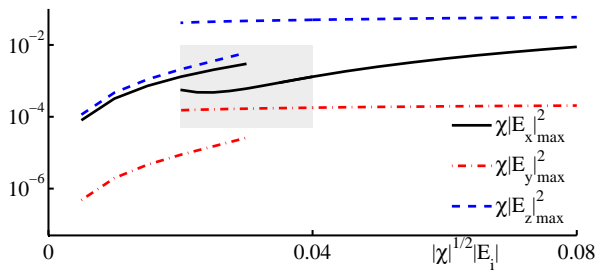}
\caption{(Color online) Semi-log plot of the maximum values, within the layered medium bulk, of the normalized squared field components as a function
of the normalized input field amplitude obtained for the full-wave evaluation of the slab transmissivity reported in Fig.5(a).}
\end{figure}

In order to discuss the feasibility of the above predicted two-parameters multistability we have performed 3D full-wave finite-element simulations
\cite{Comsol} for evaluating the transmissivity of the slab reported in Fig.1(a) in the presence of a TM radiation of free-space wavelength $\lambda
= 810$ nm. We have chosen $d_1 = 2$ nm, $d_2 = 5.25$ nm, $L = 2/k_0 = 258$ nm, $\epsilon_1 = -28.81 + 10 i$, $\epsilon_2 = 10.9 -3.8 i$, $\chi_1 =
3.16 \times 10^{-16} \: m^2/V^2$ and $\chi_2 = 0$. The parameters of medium $1$ coincide with those of silver \cite{Palikk}, characterized by a very
large nonlinear susceptibility \cite{YangGu}, with the imaginary part of the permittivity corrected by the layer size effect (since $d_1 = 2$ nm)
\cite{CaiSha} whereas medium $2$ is a linear dielectric with gain (to compensate the metal losses). The effective permittivity and nonlinear
susceptibility of the considered sample are $\epsilon = -0.054 + 0.007i$ and $\chi = 8.72 \times 10^{-17} \: m^2/V^2$. In Fig.5 we report the
comparison between the transmissivity evaluated through full-wave simulations (dotted line) and the transmissivity obtained by solving
Eqs.(\ref{Maxwell}) (solid line) in the two situations (panel (a) and (b)) corresponding to those of Fig.2 and 3. We note that good agreement exists
between the results of the two kind of simulations and, most importantly, that finite-element simulations still predicts the above discussed
multistability thus proving its robustness. The origin of the discrepancies between the transmissivities lies in the fact that a surface plasmon
resonance occurs when the TM wave impinges on the slab as reported in Fig.1(b). Therefore an electric field $y$-component ($E_y$) of plasmonic origin
arises mainly at the edges of the layers inside the medium and it is characterized by a sub-wavelength varying profile and an evanescent field tail
in vacuum which does not contribute to the power flow. Evidently, if $E_y$ is much smaller than $E_x$ and $E_z$, the homogenization theory correctly
describes the slab nonlinear behavior. In Fig.6 we report the maximum values (within the medium) attained by the three field components for $\theta =
0.17$ rad as a function of the normalized input field amplitude corresponding to the evaluated full-wave transmissivities reported in Fig.5(a). It is
evident that, outside the shaded region, $E_y$ (dot-dashed line) is much smaller than both $E_x$ and $E_z$ (solid and dashed lines respectively) and
in the corresponding regions of Fig.4(a) the agreement with the homogenization approach is very satisfactory. Within the shaded region of Fig.6,
$E_y$ is not negligible and this partially breaks the validity of the homogenization approach thus leading to the discrepancies of the
transmissivities in the range $0.02 < |\chi|^{1/2} |E_i| < 0.04$. reported in Fig.5(a).

The intensity of the incident plane wave is $I=(1/2)\sqrt{\epsilon_0 / \mu_0} |E_i|^2$ which, for the amplitude range $0.02 < |\chi|^{1/2} |E_i| <
0.04$ where multistability occurs for $\epsilon = -0.05$ (see Figs.2(a) and 4(a)) and for the above effective nonlinear susceptibility $\chi$, yields
the values $1.2 \: MW / cm^2 < I < 4.8 \: MW / cm^2$, which are intensities smaller than those normally required for observing the standard optical
bistability. However it is evident that the smaller $|\epsilon|$ the smaller the intensity required for observing the two-parameter hysteresis so
that very smaller intensity thresholds (of the order of $W/cm^2$) are very likely to be attained.

In conclusion, we have shown that a nonlinear metamaterial slab with a very small effective linear permittivity can host more than one
electromagnetic configuration compatible with the incoming field and this leads to multistability and two-parameters hysteresis steered through the
input intensity and incident angle. The two degrees of freedom potentially offers an infinite number of hysteresis loops. In addition, as a
consequence of the small value of the permittivity, hysteresis is predicted at low input intensities and this suggests that our findings can have
important applications in devising tunable and compact photonic devices for advanced optical control and memory functionalities.


\end{document}